\documentclass[11pt,a4paper]{article}

\usepackage[reqno]{amsmath} %gives new environments such as split, align etc. 
\usepackage{bbm} %for sets etc. $\mathbbm{N}$

\usepackage{array}
\usepackage{multirow}

\usepackage{graphicx}
\usepackage{float}
\usepackage{rotating}

\usepackage{feynmp}

\usepackage{a4wide}

\usepackage{simplewick}

\usepackage{xcolor}
\usepackage[numbers,sort&compress]{natbib}
\definecolor{darkblue}{rgb}{0,0,.5}
\usepackage{hyperref} 
\hypersetup{colorlinks=true, breaklinks=true, citecolor=darkblue, filecolor=darkblue, menucolor=darkblue, urlcolor=darkblue, linkcolor=darkblue,pagecolor=darkblue}
\usepackage{hypernat}

\newcommand{\Slash}[1]{#1\!\!\!/}

\begin{document}

% For feynmf-Package:
\setlength{\unitlength}{1mm}

\begin{titlepage}
\title{\vspace*{-2.0cm}
%\hfill {\small hep--ph/xxxxxx}\\[20mm]
\bf\Large
On the Quantitative Impact of the Schechter-Valle Theorem
\\[5mm]\ }

\author{
Michael Duerr$^a$\thanks{email: \tt michael.duerr@mpi-hd.mpg.de}~,~
Manfred Lindner$^a$\thanks{email: \tt manfred.lindner@mpi-hd.mpg.de}~,~
and Alexander Merle$^{ab}$\thanks{email: \tt amerle@kth.se}
\\ \\
$^a${\normalsize \it Max-Planck-Institut f\"ur Kernphysik,}\\
{\normalsize \it Postfach 10 39 80, 69029 Heidelberg, Germany}\\
\\
$^b${\normalsize \it Department of Theoretical Physics, School of Engineering Sciences,}\\
{\normalsize \it Royal Institute of Technology (KTH), AlbaNova University Center,}\\
{\normalsize \it Roslagstullsbacken 21, 106 91 Stockholm, Sweden}
}
\date{}
\maketitle

%no page numbering or other automatic things
\thispagestyle{empty}

\begin{abstract}
\noindent
We evaluate the Schechter-Valle (Black Box) theorem quantitatively by considering the most general Lorentz invariant Lagrangian consisting of point-like operators for neutrinoless double beta decay. It is well known that the Black Box operators induce Majorana neutrino masses at four-loop level. This warrants the statement that an observation of neutrinoless double beta decay guarantees the Majorana nature of neutrinos. We calculate these radiatively generated masses and find that they are many orders of magnitude smaller than the observed neutrino masses and splittings. 
%
%This implies that some lepton number violating New Physics (which may at tree-level not be related to neutrino masses) induces Black Box operators which have to explain an observed rate of neutrinoless double beta decay. This guarantees also finite Majorana neutrino masses, but the smallness of the Black Box contributions implies that other neutrino mass terms must exist, which could be of Dirac or Majorana nature, contributing to the total neutrino mass. 
%
Thus, some lepton number violating New Physics (which may at tree-level not be related to neutrino masses) may induce Black Box operators which can explain an observed rate of neutrinoless double beta decay. Although these operators guarantee finite Majorana neutrino masses, the smallness of the Black Box contributions implies that other neutrino mass terms (Dirac or Majorana) must exist.
If neutrino masses have a significant Majorana contribution then this will become the dominant part of the Black Box operator. However, neutrinos might also be predominantly Dirac particles, while other lepton number violating New Physics dominates neutrinoless double beta decay. Translating an observed rate of neutrinoless double beta decay into neutrino masses would then be completely misleading. Although the principal statement of the Schechter-Valle theorem remains valid, we conclude that the Black Box diagram itself generates radiatively only mass terms which are many orders of magnitude too small to explain neutrino masses. 
%
%Therefore, other leading contributions to neutrino masses must exist, which could be of Dirac or Majorana nature.
%
Therefore, other operators must give the leading contributions to neutrino masses, which could be of Dirac or Majorana nature.
\end{abstract}

\end{titlepage}

%%%%%%%%%%%%%%%%%%%%%%%%%%%%%%%%%%%%%%%%%%%%%%%%%%%%%%%%%%%%%%%%%%%%%%
\section{\label{sec:intro} Introduction}
%%%%%%%%%%%%%%%%%%%%%%%%%%%%%%%%%%%%%%%%%%%%%%%%%%%%%%%%%%%%%%%%%%%%%%

During the last decades neutrino flavor transitions were eventually found in oscillation experiments with atmospheric, solar, reactor, and accelerator neutrinos. These experiments have shown without doubt that neutrinos have small non-vanishing masses~\cite{Fukuda:1998mi,Ahmad:2002jz,Eguchi:2002dm} requiring some extension of the Standard Model. Different possibilities exist to generate neutrino masses, but their special quantum numbers allow them to be  Dirac or Majorana particles. This issue is directly related to the question whether lepton number is or is not a symmetry of Nature, because a Majorana mass term for the neutrino violates lepton number by two units. This question cannot be answered by standard neutrino oscillation experiments, unless one would be able to directly detect a conversion between neutrino and antineutrino (or vice versa). Therefore, other experiments have to be performed to determine the nature of neutrinos.

Unfortunately, lepton number violating processes generically have very small amplitudes, as they are usually suppressed by tiny neutrino masses. Therefore, it is very difficult to observe these processes experimentally. At the moment, the most promising attempts to find lepton number violation are the experiments searching for neutrinoless double beta decay ($0\nu\beta\beta$). Several such experiments have been performed (IGEX~\cite{Aalseth:2002rf}, Heidelberg-Moscow~\cite{KlapdorKleingrothaus:2004wj}, CUORICINO~\cite{Arnaboldi:2008ds}, NEMO~\cite{Argyriades:2008pr}, and others), so far without unambiguous detection. New second generation experiments like GERDA~\cite{Schonert:2005zn}, which has been built and is being readied for data taking,  will soon improve the sensitivity and may observe $0\nu\beta\beta$. We would like to emphasize that such a signal first of all establishes lepton number violation and that one should be a bit careful not to translate it immediately into a Majorana neutrino mass. We will therefore take a closer look at the well-known Schechter-Valle theorem~\cite{Schechter:1981bd}, which guarantees that neutrinos are Majorana particles if neutrinoless double beta decay is observed. We will determine the numerical magnitude of these guaranteed Majorana mass contributions and discuss the consequences. 

The paper is organized as follows: We first review the classic version of the Black Box theorem, as well as its extension to general lepton number and lepton flavor violating processes in Sec.~\ref{sec:bbt}. In Sec.~\ref{sec:bb_mass}, we classify all point-like operators that can contribute to $0\nu\beta\beta$ and calculate the mass corrections for two examples of operators, one of which leads to a vanishing correction while the other one generates a tiny but non-zero neutrino mass correction. We finally explain the connection of our results to chiral symmetry breaking in Sec.~\ref{sec:chiral} before we conclude in Sec.~\ref{sec:conc}. 

%%%%%%%%%%%%%%%%%%%%%%%%%%%%%%%%%%%%%%%%%%%%%%%%%%%%%%%%%%%%%%%%%%%%%%
\section{\label{sec:bbt} Schechter-Valle Theorem (Black Box Theorem)}
%%%%%%%%%%%%%%%%%%%%%%%%%%%%%%%%%%%%%%%%%%%%%%%%%%%%%%%%%%%%%%%%%%%%%%

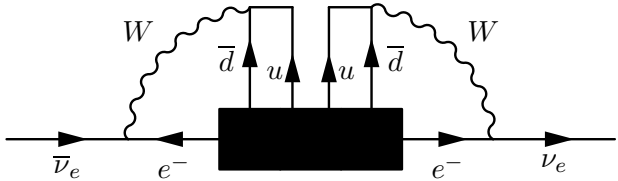
\begin{figure}
\centering
\begin{fmffile}{blackboxtheorem}
\begin{fmfgraph*}(80,25)
 \fmfleft{d1,i1,d2,d3}
 \fmfright{d4,o1,d5,d6}
 \fmf{phantom}{i1,v1,v2,v3,v4,v5,v6,v7,o1}
 \fmf{phantom}{d3,v8,v9,v10,v11,v12,v13,v14,d6}
 \fmfforce{.0w,.15h}{i1}
 \fmfforce{.2w,.15h}{v1}
 \fmfforce{.4w,.15h}{v2}
 \fmfforce{.47w,.15h}{v3}
 \fmfforce{.5w,.15h}{v4}
 \fmfforce{.53w,.15h}{v5}
 \fmfforce{.6w,.15h}{v6}
 \fmfforce{.8w,.15h}{v7}
 \fmfforce{1.0w,.15h}{o1}
 \fmfforce{.0w,.85h}{d3}
 \fmfforce{.2w,.85h}{v8}
 \fmfforce{.4w,.85h}{v9}
 \fmfforce{.47w,.85h}{v10}
 \fmfforce{.5w,.85h}{v11}
 \fmfforce{.53w,.85h}{v12}
 \fmfforce{.6w,.85h}{v13}
 \fmfforce{.8w,.85h}{v14}
 \fmfforce{1.0w,.85h}{d6}
 \fmf{fermion,label=$\overline{\nu}_e$,label.side=right}{i1,v1}
 \fmf{fermion,label=$e^-$,label.side=left}{v2,v1}
 \fmf{fermion,label=$e^-$,label.side=right}{v6,v7}
 \fmf{fermion,label=$\nu_e$,label.side=right}{v7,o1}
 \fmfv{decor.shape=square,decor.filled=full,decor.size=0.1w}{v2}
 \fmfv{decor.shape=square,decor.filled=full,decor.size=0.1w}{v4}
 \fmfv{decor.shape=square,decor.filled=full,decor.size=0.1w}{v6}
 \fmf{photon,left=0.35,label=$W$,label.side=left,tension=0}{v1,v9}
 \fmf{photon,right=0.35,label=$W$,label.side=right,tension=0}{v7,v13}
 \fmf{fermion,label=$\overline{d}$,label.side=left}{v2,v9}
 \fmf{fermion,label=$u$,label.side=left,label.dist=0.015w}{v3,v10}
 \fmf{fermion,label=$u$,label.dist=0.015w}{v5,v12}
 \fmf{fermion,label=$\overline{d}$}{v6,v13}
 \fmf{plain}{v9,v10}
 \fmf{plain}{v12,v13}
\end{fmfgraph*}
\end{fmffile}
\caption{\label{fig:blackboxtheorem} Contribution of the Black Box operator to the Majorana neutrino mass \cite{Schechter:1981bd}.}
\end{figure}

The  Schechter-Valle theorem (Black Box theorem)~\cite{Schechter:1981bd} relates the effective neutrinoless double beta decay operator to a non-zero effective Majorana electron neutrino mass. The underlying argument can easily be explained using Fig.~\ref{fig:blackboxtheorem}.
Effectively, neutrinoless double beta decay may be seen as a ``scattering'' process ($0 \rightarrow \overline{d}\,\overline{d}\,u\,u\,e\,e$) because, under the assumption that the weak interaction is described by a local gauge theory, crossing symmetry holds. One only has to additionally assume that~\cite{Takasugi:1984xr}
\begin{enumerate}
 \item the $u$ and $d$ quarks and the electron are massive, and
 \item the standard left-handed interaction $\left( \overline{\nu_{eL}} \gamma_{\mu} e_L + \overline{u_L} \gamma_{\mu} d_L \right) W^\mu$ exists.
\end{enumerate}
Then, it is possible to draw the diagram in Fig.~\ref{fig:blackboxtheorem}, so that neutrinoless double beta decay induces a non-zero effective Majorana mass for the electron neutrino, no matter which is the underlying mechanism of the decay. The Black Box is nothing but an effective operator for neutrinoless double beta decay which arises from some underlying New Physics. The first assumption is necessary to ensure that two identical neutrinos are created. This can be seen in the following way~\cite{Takasugi:1984xr}: We do not know anything about the chirality of the electrons and quarks produced by neutrinoless double beta decay. However, this assumption guarantees that we can make the particles running in the loops in Fig.~\ref{fig:blackboxtheorem} left-handed, by mass insertion if necessary. Thus the standard left-handed interaction from the second assumption produces the same type of neutrino at both vertices. Otherwise it would be possible that a neutrino and an antineutrino are created, which would give a Dirac mass term.

Note, however, that the diagram in Fig.~\ref{fig:blackboxtheorem} is certainly not the only one that generates a non-zero effective Majorana mass for the electron neutrino. Other tree and loop diagrams exist and in addition the physical neutrino masses depend also on Dirac mass terms. Furthermore, there may even be cancellations between different Majorana contributions which are induced by the Black Box diagram(s). This may appear as a fine-tuning, but the observed fermion mass patterns suggest that symmetries which explain these 
patterns may exist, and such symmetries could also lead to non-trivial cancellations. Taking into account this possibility of cancellations, Takasugi~\cite{Takasugi:1984xr} and Nieves~\cite{Nieves:1984sn} improved the argument of Schechter and Valle~\cite{Schechter:1981bd}, and showed that there cannot be a continuous or discrete symmetry protecting a vanishing Majorana mass to all orders in perturbation theory. We will follow the arguments of Takasugi~\cite{Takasugi:1984xr} here. He assumed an unbroken discrete symmetry protecting the Majorana neutrino mass (the $\eta$'s are global phase factors):
\begin{equation}
 \nu_{eL} \rightarrow \eta_{\nu} \nu_{eL}, \quad e_L \rightarrow \eta_e e_L, \quad  q_L \rightarrow \eta_q q_L \; (q=u,d) , \quad  W^{+\mu}_L \rightarrow \eta_W W^{+\mu}_L \, .
\end{equation}
To forbid the Majorana mass term, we need to have
\begin{equation}
 \eta_{\nu}^2 \neq 1\, ,
 \label{eq:tak1}
\end{equation}
and the invariance of the left-handed interaction requires
\begin{equation}
 \eta_{\nu}^{\ast} \eta_e \eta_W = \eta_u^{\ast} \eta_d \eta_W = 1\, .
 \label{eq:tak2}
\end{equation}
However, the existence of $0\nu \beta\beta$ (that is, the process $d_L+d_L \rightarrow u_L+u_L+e_L+e_L$) implies
\begin{equation}
 \eta_u^2 \eta_d^{\ast 2} \eta_e^2 = 1\, .
 \label{eq:tak3}
\end{equation}

It is easy to see that Eqs.~\eqref{eq:tak1}, \eqref{eq:tak2}, and~\eqref{eq:tak3} cannot be solved simultaneously. Thus, if the Majorana mass term is forbidden by an unbroken discrete symmetry, there will be no neutrinoless double beta decay. On the other hand, if neutrinoless double beta decay exists, there cannot be a symmetry protecting the Majorana mass and the term will be induced (the possibility of accidental cancellation to all orders in perturbation theory is not excluded, but appears to be very unlikely).

The implications are quite strong, and this becomes particularly clear when we think of contributions to neutrinoless double beta decay from New Physics beyond the Standard Model. Besides the usual mass mechanism involving the internal exchange of a virtual Majorana neutrino, many models of New Physics allow neutrinoless double beta decay also via other lepton number violating operators. Thus the theorem states that it is not possible to construct models with massless neutrinos where $0\nu\beta\beta$ occurs.

However, it cannot be overemphasized that the theorem so far is only a qualitative statement and does not say anything about the size of the induced Majorana neutrino mass. A quantitative determination of the mass generated for different realizations of the Black Box by the diagram in Fig.~\ref{fig:blackboxtheorem} is therefore a very interesting question. We will do such a calculation for different operators using an effective field theory approach independent of the underlying model in Sec.~\ref{sec:bb_mass}. The results will be interesting, as they clearly show the limitations of the Black Box theorem.

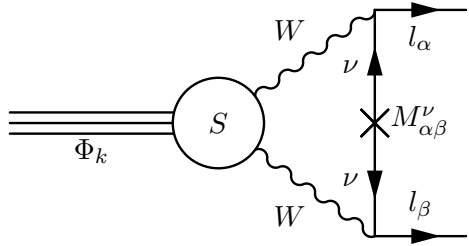
\begin{figure}[t]
\begin{center}
\begin{fmffile}{lnv1}
\begin{fmfgraph*}(60,30)
\fmfstraight
\fmfleft{i2,i1,i3}
\fmfright{o1,g1,o2}
\fmf{phantom}{i3,a,b,o2}
\fmf{phantom}{i2,c,d,o1}
\fmf{fermion,label=$l_\alpha$,label.side=right}{b,o2}
\fmf{fermion,label=$l_\beta$,label.side=left}{d,o1}
\fmffreeze
\fmf{phantom}{i1,v1,b}
\fmf{phantom}{i1,v1,d}
\fmf{phantom}{b,v4,d}
\fmf{plain,label=$\Phi_k$}{i1,v1}
\fmf{photon,label=$W$,label.side=left}{v1,b}
\fmf{photon,label=$W$,label.side=right}{v1,d}
\fmf{fermion,tension=0,label=$\nu$,label.side=left}{v4,b}
\fmf{fermion,tension=0,label=$\nu$,label.side=right}{v4,d}
\fmffreeze
\fmfi{plain}{vpath (__i1,__v1) shifted (thick*(0,2))}
\fmfi{plain}{vpath (__i1,__v1) shifted (thick*(0,-2))}
\fmfv{decor.shape=circle,decor.filled=empty,decor.size=0.2w,l=$S$,label.dist=0}{v1}
\fmfv{decoration.shape=cross,label=$M_{\alpha \beta}^\nu$}{v4}
\end{fmfgraph*}
\end{fmffile}
\caption{\label{fig:lnv1} Contribution of the $M_{\alpha \beta}^\nu$ entry of the Majorana neutrino mass matrix to the effective lepton number and lepton flavor violating vertex $\Gamma_{\alpha\beta}$.}
 \end{center}
\end{figure}

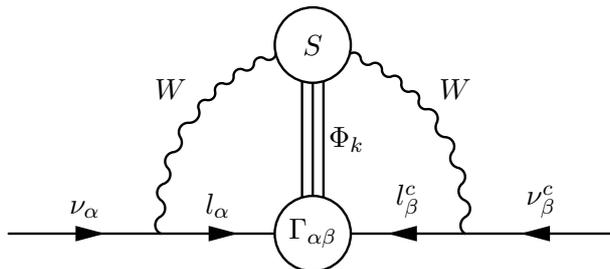
\begin{figure}
 \begin{center}
\begin{fmffile}{massterm1}
\begin{fmfgraph*}(100,25)
\fmfleft{i1,i2}
\fmfright{o1,o2}
\fmf{phantom}{i1,v1,v2,v3,o1}
\fmf{phantom}{i2,v4,v5,v6,o2}
\fmffreeze
\fmf{fermion,label=$\nu_\alpha$,label.side=left}{i1,v1}
\fmf{fermion,label=$l_\alpha$,label.side=left}{v1,v2}
\fmf{fermion,label=$l_\beta^c$}{v3,v2}
\fmf{fermion,label=$\nu_\beta^c$}{o1,v3}
\fmf{boson,label=$W$,left=0.4}{v1,v5,v3}
\fmf{plain,label=$\Phi_k$}{v2,v5}
\fmffreeze
\fmfi{plain}{vpath (__v2,__v5) shifted (thick*(-2,0))}
\fmfi{plain}{vpath (__v2,__v5) shifted (thick*(2,0))}
\fmfv{decor.shape=circle,decor.filled=empty,decor.size=0.1w,l=$S$,label.dist=0}{v5}
\fmfv{decor.shape=circle,decor.filled=empty,decor.size=0.1w,l=$\Gamma_{\alpha\beta}$,label.dist=0}{v2}
\end{fmfgraph*}
\end{fmffile}
\caption{\label{fig:massterm1} Contribution of the effective lepton number and lepton flavor violating vertex $\Gamma_{\alpha\beta}$ to the $M_{\alpha \beta}^\nu$ entry of the Majorana neutrino mass matrix.}
 \end{center}
\end{figure}

Note that there exists also an extended Black Box theorem \cite{Hirsch:2006yk} which takes the three-generation Majorana neutrino mass matrix into account and includes the fact that $\nu_e$, $\nu_{\mu}$, and $\nu_{\tau}$ mix. Moreover, the theorem has been extended to arbitrary lepton number and lepton flavor violating processes. It was also found that there exists a general set of one-to-one correspondence relations between the effective operators generating these processes and the elements of the neutrino mass matrix. In particular, $\Delta L = 2$ processes described by $\Phi_k \rightarrow l_\alpha l_\beta$ conserving baryon number have been discussed, which means that lepton number violation manifests itself via two external charged leptons in the final state. $\Phi_k$ is a set of external particles with $B=L=0$ and electrical charge $Q= -2$. Note that, for a lepton number violating process with the same flavor states $l_\alpha$ and $l_\beta$, different sets of external particles $\Phi_k$ are possible. With a symmetry argument, similar to the one used before to prove the classic Black Box theorem, it has also been shown that the following relation between the effective lepton number and lepton flavor violating vertex $\Gamma_{\alpha\beta}$ and the entry $M_{\alpha \beta}^\nu$ of the Majorana neutrino mass matrix exists:
\begin{equation}
 M_{\alpha \beta}^\nu = 0 \Leftrightarrow \Gamma_{\alpha\beta}=0\, .
\end{equation}
The corresponding diagrams are shown in Figs.~\ref{fig:lnv1} and~\ref{fig:massterm1}, which have already been discussed in \cite{Hirsch:2006yk}.

There has been some more work on the Black Box theorem: Hirsch et al.\ proved a supersymmetric version of it~\cite{Hirsch:1997dm,Hirsch:1997vz}. This extends the relation of the theorem to the lepton number violating scalar neutrino mass. We will not discuss that work further, as it is not of great importance for the discussion presented here.

%%%%%%%%%%%%%%%%%%%%%%%%%%%%%%%%%%%%%%%%%%%%%%%%%%%%%%%%%%%%%%%%%%%%%%%%%
\section{\label{sec:bb_mass} The Neutrino Mass generated by the Black Box Diagram}
%%%%%%%%%%%%%%%%%%%%%%%%%%%%%%%%%%%%%%%%%%%%%%%%%%%%%%%%%%%%%%%%%%%%%%%%%

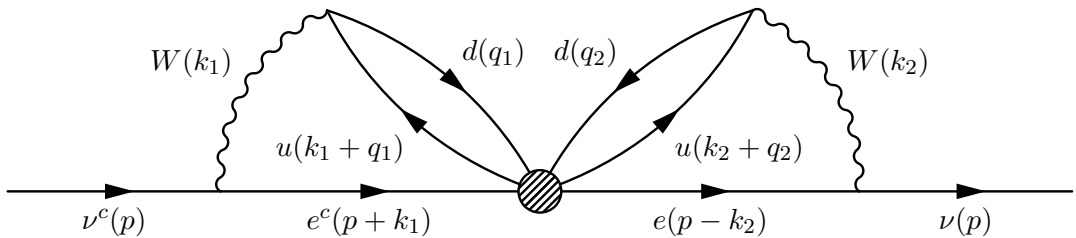
\begin{figure}[b]
\centering
\begin{fmffile}{blackboxmass}
\begin{fmfgraph*}(140,30)
\fmfleft{i}
\fmfright{o}
\fmf{phantom}{i,v1,v2,v3,o}
\fmf{fermion,label=$\nu^c(p)$}{i,v1}
\fmf{fermion,label=$e^c(p+k_1)$}{v1,v2}
\fmf{fermion,label=$e(p-k_2)$}{v2,v3}
\fmf{fermion,label=$\nu(p)$}{v3,o}
\fmf{phantom}{v2,v4,v1}
\fmf{phantom}{v2,v5,v3}
\fmf{photon,label=$W(k_1)$,right=.3}{v4,v1}
\fmf{photon,label=$W(k_2)$,left=.3}{v5,v3}
\fmf{fermion,left=.2,label=$d(q_1)$}{v4,v2}
\fmf{fermion,left=.2,label=$u(k_1+q_1)$}{v2,v4}
\fmf{fermion,right=.2,label=$u(k_2+q_2)$}{v2,v5}
\fmf{fermion,right=.2,label=$d(q_2)$}{v5,v2}
\fmfforce{.3w,.9h}{v4}
\fmfforce{.2w,.1h}{v1}
\fmfforce{.5w,.1h}{v2}
\fmfforce{.8w,.1h}{v3}
\fmfforce{.7w,.9h}{v5}
\fmfforce{.0w,.1h}{i}
\fmfforce{1w,.1h}{o}
\fmfv{decor.shape=circle,decor.filled=shaded,decor.size=8thick}{v2}
\end{fmfgraph*}
\end{fmffile}
\caption{\label{fig:blackboxmass} The ``Butterfly'' diagram to be calculated in Sec.~\ref{sec:bb_mass}.}
\end{figure}

Now we want to calculate the mass correction (``Butterfly'') which is induced by the Black Box diagram shown in Fig.~\ref{fig:blackboxmass}. We will first give a general parameterization of $0\nu\beta\beta$ in terms of effective point-like operators, and then calculate the diagram for two examples of such operators in dimensional regularization. We will find that the Schechter-Valle theorem is not necessarily as useful as it might seem: The mass that is generated by the diagram in Fig.~\ref{fig:blackboxmass} is many orders of magnitude smaller than what is expected for the neutrino mass. Moreover, there exist operators generating $0\nu\beta\beta$, but giving zero contribution to the Majorana neutrino mass via this particular diagram. Of course other diagrams, although more strongly suppressed, may give a non-zero contribution also for these operators.

%%%%%%%%%%%%%%%%%%%%%%%%%%%%%%%%%%%%%%%%%%%%%%%%%%%%%%%%%%%%%%%%%%%%%%%%%
\subsection{Parameterization of Neutrinoless Double Beta Decay}
%%%%%%%%%%%%%%%%%%%%%%%%%%%%%%%%%%%%%%%%%%%%%%%%%%%%%%%%%%%%%%%%%%%%%%%%%

The most general Lorentz invariant Lagrangian which contributes to $0\nu\beta\beta$ via point-like operators has the following form~\cite{Pas:2000vn}:
\begin{equation}
 \mathcal{L}=\frac{G_F^2}{2}m_p^{-1} \left( \epsilon_1 JJj + \epsilon_2 J^{\mu\nu} J_{\mu\nu} j + \epsilon_3 J^\mu J_\mu j + \epsilon_4 J^\mu J_{\mu\nu} j^\nu + \epsilon_5 J^\mu J j_\mu \right)\, ,
 \label{eq:L_0nbb}
\end{equation}
where $G_F$ is the Fermi coupling constant, and $m_p$ is the proton mass. The hadronic currents required are given by
\begin{equation}
 J = \overline{u}\left( 1 \pm \gamma_5 \right) d , \; J^\mu = \overline{u} \gamma^\mu \left( 1 \pm \gamma_5 \right) d, \; J^{\mu\nu} = \overline{u} \frac{i}{2} \left[\gamma^\mu, \gamma^\nu\right] \left( 1 \pm \gamma_5 \right) d\, ,
 \label{eq:operators_1}
\end{equation}
and the leptonic currents are
\begin{equation}
 j = \overline{e}\left( 1 \pm \gamma_5 \right) e^c , \; j^\mu = \overline{e} \gamma^\mu \left( 1 \pm \gamma_5 \right) e^c\, .
 \label{eq:operators_2}
\end{equation}
In the remainder of this paper we will often use the abbreviations
\begin{equation}
 P_L = \frac{1}{2} (1-\gamma_5) \text{~and~} P_R = \frac{1}{2} (1+\gamma_5)
\end{equation}
and write
\begin{equation}
 j_L = \overline{e}\left( 1 - \gamma_5 \right) e^c, \text{~etc.}
\end{equation}

For all currents, different chirality structures are permitted (cf.\ Sec.~\ref{sec:chiral}). Note that in Eq.~\eqref{eq:L_0nbb} we have already dropped the following terms which are Lorentz invariant, too:
\begin{equation}
 \mathcal{L}'=\frac{G_F^2}{2}m_p^{-1} \left(  \epsilon_6 J^\mu J^\nu j_{\mu\nu} + \epsilon_7 J J^{\mu\nu} j_{\mu\nu} + \epsilon_8 J_{\mu\alpha} J^{\nu\alpha} j_\nu^\mu \right) \, ,
 \label{eq:Lprime_0nbb}
\end{equation}
where the leptonic tensor current is given by
\begin{equation}
 j^{\mu\nu} = \overline{e} \frac{i}{2} \left[\gamma^\mu, \gamma^\nu\right] \left( 1 \pm \gamma_5 \right) e^c\, .
\end{equation}
These terms had been included in the Lagrangian density in~\cite{Pas:2000vn}, but were neglected in the final analysis, as the authors worked in the $s$-wave approximation where these contributions vanish. In~\cite{Prezeau:2003xn}, it was pointed out that all operators proportional to
\begin{equation}
 \overline{e}\gamma_\mu e^c,\; \overline{e}\frac{i}{2}\left[\gamma_\mu,\gamma_\nu\right] e^c, \text{~and~} \overline{e}\gamma_5 \frac{i}{2}\left[\gamma_\mu,\gamma_\nu\right] e^c
\end{equation}
vanish identically because of the electron fields being Grassmann numbers. Therefore, the terms in Eq.~\eqref{eq:Lprime_0nbb} are not relevant for neutrinoless double beta decay.

We may have to distinguish different chiralities of the currents involved, as in some cases the decay rate depends on them. Therefore, the authors of~\cite{Pas:2000vn} used the parameters $\epsilon_i^{xyz}$ with $x,y,z = L/R$ to define the chiralities of the hadronic and leptonic currents in order of their appearance in Eq.~\eqref{eq:L_0nbb}. A suppressed chirality index indicates that it is not necessary to distinguish different chiralities. The limits obtained in~\cite{Pas:2000vn} for the parameters $\epsilon_i$ are given in Tab.~\ref{tab:epsilons}. We re-calculated the values of interest for us, using more recent results for the nuclear matrix elements ($\mathcal{M}_{GT,N} = 1.70\times 10^{-1}$ and $\mathcal{M}_{F,N} = 6.80\times 10^{-2}$) calculated in the Self-consistent Renormalized Quasiparticle Random Phase Approximation (SRQRPA)~\cite{Simkovic:2010ka,Simkovic:2011}. Moreover, we have used $g_A = 1.25$ and the half-life limit $T_{1/2}^{0\nu} > 1.8\times 10^{25} \,\mathrm{y}$ obtained in the Heidelberg-Moscow experiment, as used in~\cite{Pas:2000vn}. Note that, for $\epsilon_3$, in addition to the updated values of the nuclear matrix elements, our cross-check of the bound from Ref.~\cite{Pas:2000vn} resulted in a number that was smaller by a factor of 2~\cite{Bergstrom:2011dt}.

Let us mention that of course long range contributions to $0\nu\beta\beta$---not proportional to the Majorana neutrino mass---may exist. A calculation of these diagrams, however, is not expected to add new insights.

\renewcommand{\arraystretch}{1.5}
\begin{table}
\centering
 \begin{tabular}{ccccccc}\hline
 & $|\epsilon_1|$ & $|\epsilon_2|$ & $|\epsilon_3^{LLz}|$, $|\epsilon_3^{RRz}|$ & $|\epsilon_3^{LRz}|$, $|\epsilon_3^{RLz}|$ & $|\epsilon_4|$ & $|\epsilon_5|$ \\ \hline\hline
from \cite{Pas:2000vn}&$3\times 10^{-7}$ & $2\times 10^{-9}$ & $4\times 10^{-8}$ & $1\times 10^{-8}$ & $2\times 10^{-8}$ & $2\times 10^{-7}$ \\ 
our calc.\ & $2.0\times 10^{-7}$ & & $1.5\times 10^{-8}$ &  & & \\ \hline
 \end{tabular}
\caption{\label{tab:epsilons} Upper bounds for the coupling parameters in Eq.~\eqref{eq:L_0nbb}. They were evaluated ``on axis,'' meaning that all other contributions were set to zero to extract the limits on one of the parameters. Different chiralities in the hadronic currents lead to different values only in the case of $\epsilon_3$.}
\end{table}

%%%%%%%%%%%%%%%%%%%%%%%%%%%%%%%%%%%%%%%%%%%%%%%%%%%%%%%%%%%%%%%%%%%%%%%%%
\subsection{Vertices and Propagators}
%%%%%%%%%%%%%%%%%%%%%%%%%%%%%%%%%%%%%%%%%%%%%%%%%%%%%%%%%%%%%%%%%%%%%%%%%

The Butterfly diagram we have to calculate (Fig.~\ref{fig:blackboxmass}) is a non-standard one. At the effective vertex, lepton number is violated and two electrons are produced, which then leads to two outgoing neutrinos. We want to have a continuous fermion line in our diagram, so we have to rewrite some of the outgoing fields as incoming charge conjugate fields. The leptonic part of the standard electroweak vertex is given by 
\begin{equation}
 \frac{g}{\sqrt{2}} \left(\overline{e} \gamma^\mu P_L \nu  W_\mu + \mathrm{H.c.}\right)\, .
\end{equation}
We are interested in the Hermitian conjugate part, which is responsible for the annihilation of an incoming electron and the creation of an outgoing neutrino, and can be written as 
\begin{equation}
 \left( \overline{e} \gamma^\mu P_L \nu W_\mu \right)^\dagger = \overline{\nu} \gamma^\mu P_L e W_\mu\,  .
\end{equation}
To rewrite it in terms of incoming charge conjugate fields, we have to transpose this expression. Using $C \gamma^{\mu T} C^{-1} = - \gamma^\mu$, $C \gamma_5^T C^{-1} = \gamma_5$, and $\overline{\psi^c} = -\psi^T C^{-1}$ we obtain
\begin{equation}
 \left(\overline{\nu} \gamma^\mu P_L e \right)^T =\overline{e^c} \gamma^\mu P_R \nu^c\, .
 \label{eq:gamma_rel}
\end{equation}
We additionally have to calculate the propagator of the charge conjugate electron fields:
\begin{equation}
\contraction{}{e^c}{(y)}{\overline{e^c}}
e^c(y) \overline{e^c}(x)\, .
\label{eq:propcharge}
\end{equation}
Starting from the usual electron propagator
\begin{equation}
 \contraction{}{e}{(x)}{\overline{e}}
e(x) \overline{e}(y) = i S_F(x-y)
\end{equation}
with
\begin{equation}
 S_F(x-y) = \int \frac{d^4p}{(2\pi)^4} \frac{\Slash{p}+m}{p^2-m^2+i\epsilon} e^{-ip(x-y)}\, ,
 \label{eq:prop}
\end{equation}
it is easy to see that
\begin{equation}
 \contraction{}{\overline{e}^T}{(y)}{e^T}
\overline{e}^T(y)e^T(x) = i S_F^T(x-y)\, .
\end{equation}
If we now rewrite the expression in Eq.~\eqref{eq:propcharge} with the help of
\begin{equation}
 e^c(y) = C\overline{e}^T(y)
\end{equation}
and the relation
\begin{equation}
 C S_F^T(x-y) C^{-1} = S_F(y-x)\, ,
\end{equation}
which follows directly from the relations used to arrive at Eq.~\eqref{eq:gamma_rel}, we obtain
\begin{equation}
 \contraction{}{e^c}{(y)}{\overline{e^c}}
e^c(y) \overline{e^c}(x) = S_F(y-x)\, . 
\end{equation}
In the remainder of this section, we will work in momentum space. Therefore, a look at Eq.~\eqref{eq:prop} reveals that the change of sign in the argument of $S_F$ changes the sign of the momentum $\Slash{p}$. We thus have in momentum space
\begin{equation}
  \contraction{}{e^c}{(y)}{\overline{e^c}}
e^c(y) \overline{e^c}(x) \sim i \frac{-\Slash{p}+m}{p^2-m^2+i\epsilon}\, .
\end{equation}

%%%%%%%%%%%%%%%%%%%%%%%%%%%%%%%%%%%%%%%%%%%%%%%%%%%%%%%%%%%%%%%%%%%%%%%%%
\subsection{Decay mediated by the Operator \texorpdfstring{$J_LJ_Lj_L$}{JLJLjL}}
%%%%%%%%%%%%%%%%%%%%%%%%%%%%%%%%%%%%%%%%%%%%%%%%%%%%%%%%%%%%%%%%%%%%%%%%%

Our first explicit example will be the operator 
\begin{equation}
 J_LJ_Lj_L = 8 \overline{u} P_L d \, \overline{u} P_L d \, \overline{e} P_L e^c \, .
\end{equation}
To calculate the diagram in Fig.~\ref{fig:blackboxmass}, we have to write down the matrix element. Let us define the weak leptonic current
\begin{equation}
 j_l^\alpha = \overline{\nu} \gamma^\alpha P_L e
\end{equation}
and the weak hadronic current
\begin{equation}
 J_h^\mu = \overline{d} \gamma^\mu P_L u\, .
\end{equation}
We then have to find all possible contractions in 
\begin{equation}
 \langle f | J_LJ_Lj_L j_l^{\nu T} W_\nu J_h^\mu W_\mu j_l^\alpha W_\alpha J_h^\beta W_\beta | i\rangle \, ,
\end{equation}
where $| i\rangle$ and $| f \rangle$ denote initial and final state, respectively. It is easy to see that all contractions lead to the same diagram given in Fig.~\ref{fig:blackboxmass}. We find
\begin{equation}
\contraction{\langle }{f}{|}{\overline{\nu}}
\contraction{\langle f |\overline{\nu} \gamma^\alpha P_L }{e}{}{\overline{e}}
\contraction{\langle f |\overline{\nu} \gamma^\alpha P_L e \overline{e} P_L }{e^c}{}{\overline{e^c}}
\contraction{\langle f |\overline{\nu} \gamma^\alpha P_L e \overline{e} P_L e^c \overline{e^c} \gamma^\nu P_R }{\nu^c}{ \overline{u} P_L d \overline{d} \gamma^\mu P_L u \overline{u} P_L d\overline{d} \gamma^\beta P_L u W_\nu W_\mu W_\alpha W_\beta |}{i}
\bcontraction[2ex]{ \langle f |\overline{\nu} \gamma^\alpha P_L e \overline{e} P_L e^c \overline{e^c} \gamma^\nu P_R \nu^c }{\overline{u}}{P_L d \overline{d} \gamma^\mu P_L }{u}
\bcontraction{\langle f |\overline{\nu} \gamma^\alpha P_L e \overline{e} P_L e^c \overline{e^c} \gamma^\nu P_R \nu^c \overline{u} P_L }{d}{}{\overline{d}}
\bcontraction[2ex]{ \langle f |\overline{\nu} \gamma^\alpha P_L e \overline{e} P_L e^c \overline{e^c} \gamma^\nu P_R \nu^c \overline{u} P_L d \overline{d} \gamma^\mu P_L u }{\overline{u}}{P_L d\overline{d} \gamma^\beta P_L }{u}
\bcontraction{\langle f |\overline{\nu} \gamma^\alpha P_L e \overline{e} P_L e^c \overline{e^c} \gamma^\nu P_R \nu^c \overline{u} P_L d \overline{d} \gamma^\mu P_L u \overline{u} P_L }{d}{}{\overline{d}}
\bcontraction{\langle f |\overline{\nu} \gamma^\alpha P_L e \overline{e} P_L e^c \overline{e^c} \gamma^\nu P_R \nu^c \overline{u} P_L d \overline{d} \gamma^\mu P_L u \overline{u} P_L d\overline{d} \gamma^\beta P_L u }{W_\nu}{}{W_\mu}
\bcontraction{\langle f |\overline{\nu} \gamma^\alpha P_L e \overline{e} P_L e^c \overline{e^c} \gamma^\nu P_R \nu^c \overline{u} P_L d \overline{d} \gamma^\mu P_L u \overline{u} P_L d\overline{d} \gamma^\beta P_L u W_\nu W_\mu }{W_\alpha}{}{W_\beta}
 \langle f |\overline{\nu} \gamma^\alpha P_L e \overline{e} P_L e^c \overline{e^c} \gamma^\nu P_R \nu^c \overline{u} P_L d \overline{d} \gamma^\mu P_L u \overline{u} P_L d\overline{d} \gamma^\beta P_L u W_\nu W_\mu W_\alpha W_\beta |i\rangle \, .
\end{equation}
Different contractions lead to $8$ diagrams (there are 4 ways to contract the quark fields, and independently there are 2 ways to contract the $W$ boson fields). Using the information from the last subsection and including all necessary factors, we can directly write down the matrix element of the process:
\begin{multline}\label{eq:diagram1}
% \begin{split}
 \frac{8g^4 G_F^2 \epsilon_1}{m_p} \int d\tilde{k}_1 d\tilde{k}_2 d\tilde{q}_1 d\tilde{q}_2 \frac{-i g_{\alpha\beta}}{(k_1^2-M_W^2)}\frac{-i g_{\mu\nu}}{(k_2^2-M_W^2)}\\
\mathrm{Tr} \left(i \frac{\Slash{k}_1+\Slash{q}_1+m_u}{(k_1+q_1)^2-m_u^2} P_L i \frac{\Slash{q}_1+m_d}{q_1^2-m_d^2} \gamma^\beta P_L \right) \mathrm{Tr} \left(i \frac{\Slash{k}_2+\Slash{q}_2+m_u}{(k_2+q_2)^2-m_u^2} P_L i \frac{\Slash{q}_2+m_d}{q_2^2-m_d^2} \gamma^\mu P_L \right) \\
\overline{\nu}(p) \gamma^\alpha P_L i \frac{\Slash{p}-\Slash{k}_2+m_e}{(p-k_2)^2-m_e^2} P_L i \frac{-\Slash{p}-\Slash{k}_1+m_e}{(p+k_1)^2-m_e^2} \gamma^\nu P_R \nu^c(p) \, ,
% \end{split}
\end{multline}
where we have used the short-hand notation $d\tilde{k} = d^4 k/(2\pi)^4$. The traces can be calculated in the standard manner, yielding
\begin{equation}
\mathrm{Tr} \left(i \frac{\Slash{k}_1+\Slash{q}_1+m_u}{(k_1+q_1)^2-m_u^2} P_L i \frac{\Slash{q}_1+m_d}{q_1^2-m_d^2} \gamma^\beta P_L \right) = -\frac{2 (q_1)^\beta m_u}{[(k_1+q_1)^2-m_u^2] (q_1^2-m_d^2)}
\end{equation}
and
\begin{equation}
 \mathrm{Tr} \left(i \frac{\Slash{k}_2+\Slash{q}_2+m_u}{(k_2+q_2)^2-m_u^2} P_L i \frac{\Slash{q}_2+m_d}{q_2^2-m_d^2} \gamma^\mu P_L \right) = - \frac{2 (q_2)^\mu m_u}{[(k_2+q_2)^2-m_u^2] (q_2^2-m_d^2)}\, .
\end{equation}
We may simplify the last line of Eq.~\eqref{eq:diagram1} by moving the $P_L$'s to the right. We thus project out the terms containing $m_e$. The final expression for the self-energy of the neutrino at this order in perturbation theory is
\begin{equation}
 \Sigma(p) = \frac{32 g^4 G_F^2 \epsilon_1 m_u^2 m_e^2 }{m_p}  \mathcal{I}\, ,
 \label{eq:diagram1_intermed}
\end{equation}
where the integral $\mathcal{I}$ is given by
\begin{equation}
 \mathcal{I} = \mathcal{I}_1 \times \mathcal{I}_2\, ,
\end{equation}

\noindent with
\begin{equation}
 \mathcal{I}_1 = \int d\tilde{k}_1 d\tilde{q}_1 \frac{\Slash{q}_1}{[(p+k_1)^2-m_e^2][(k_1+q_1)^2-m_u^2](q_1^2-m_d^2)(k_1^2-M_W^2)}
\end{equation}

\noindent and
\begin{equation}
 \mathcal{I}_2 = \int d\tilde{k}_2 d\tilde{q}_2 \frac{\Slash{q}_2}{[(p-k_2)^2-m_e^2][(k_2+q_2)^2-m_u^2](q_2^2-m_d^2)(k_2^2-M_W^2)}\, .
\end{equation}

\noindent Integrals of this kind typically arise in two-loop calculations, which is not surprising: Our four-loop diagram factorizes because of the point-like nature of the corresponding operators, and it basically consists of two two-loop diagrams. These integrals may be calculated analytically in terms of Spence functions, for which extensive literature exists. Such calculations have, for example, been done in~\cite{vanderBij:1983bw,Ghinculov:1994sd,Ghinculov:1997pd}, in the framework of Higgs physics. For the moment, it suffices to know that the integrals are well behaved. To directly extract the dependence on masses and momenta, we will go another way:
We can calculate $\mathcal{I}_1$ and $\mathcal{I}_2$ separately with the help of Feynman parameters and dimensional regularization. For example, let us calculate the integral
\begin{equation}
 \mathcal{I}_1 = \int d\tilde{q}_1 \frac{\Slash{q}_1}{(q_1^2-m_d^2)} \mathcal{I}_1^\prime,
 \label{eq:I-integral}
\end{equation}
where
\begin{equation}
 \mathcal{I}_1^\prime = \int d\tilde{k}_1 \frac{1}{[(p+k_1)^2-m_e^2][(k_1+q_1)^2-m_u^2] (k_1^2-M_W^2)}\, .
\end{equation}
Using Feynman parameterization, we can rewrite
\begin{multline}
 \mathcal{I}_1^\prime = \int_0^1 dx \int_0^{1-x} dy \\
\times \int d\tilde{k}_1 \frac{2}{\left\{x[(p+k_1)^2-m_e^2]+y[(k_1+q_1)^2-m_u^2]+(1-x-y) (k_1^2-M_W^2)\right\}^3}\, .
\end{multline}
The denominator can be rewritten as $\left\{ [k_1+(xp+yq_1)]^2 -\Delta_1^\prime \right\}^3$, with
\begin{equation}
 \Delta_1^\prime = (xp+yq_1)^2-xp^2-yq_1^2+(1-x-y)M_W^2\, .
\end{equation}
Here, we have neglected $m_e^2$ and $m_u^2$ in comparison to $M_W^2$. Performing the substitution $ k_1 \rightarrow k_1-(xp+yq_1)$, plugging the result into Eq.~\eqref{eq:I-integral}, and introducing a third Feynman parameter, we obtain
\begin{multline}
 \mathcal{I}_1 = \frac{-i}{16\pi^2}\int_0^1 dx \int_0^{1-x} dy \int_0^1 dz \\ \times \int d\tilde{q}_1\frac{\Slash{q}_1}{\left\{(1-z)(q_1^2-m_d^2)+ z [(xp+yq_1)^2-xp^2-yq_1^2+(1-x-y)M_W^2]\right\}^2}\, .
\end{multline}
After completing the square and neglecting $m_d \ll M_W$, we can write the denominator as
\begin{equation}
 (1-z+zy^2-zy) \left[\left(q_1+\frac{xyz}{1-z+zy^2-zy}p\right)^2-\Delta_1(x,y,z)\right]\, , 
\end{equation}
where
\begin{multline}\label{eq:delta1}
 \Delta_1(x,y,z) =  \left( \frac{xyz}{(1-z+zy^2-zy)} p\right)^2 \\
- \frac{1}{(1-z+zy^2-zy)} \left[zx^2p^2-zxp^2+(1-x-y)zM_W^2-m_d^2\right]\, .
\end{multline}
Plugging this result back into the expression for $\mathcal{I}_1$ in Eq.~\eqref{eq:I-integral}, performing one more linear substitution, dropping all terms linear in $q_1$ (as they vanish in the integration), and finally performing the integral over $q_1$, we arrive at
\begin{multline}
 \mathcal{I}_1 = \frac{\Slash{p}}{16\pi^2}\int_0^1 dx \int_0^{1-x} dy  \\
 \times \int_0^1 dz\frac{xyz}{(1-z+zy^2-zy)^3} \left(\frac{2}{\epsilon} -\log \Delta_1 -\gamma+\log 4\pi + \mathcal{O}(\epsilon) \right)\, .
\end{multline}
In this expression, $\epsilon = 4-d$ and $\gamma \approx 0.5772$ is the Euler-Mascheroni constant. 
Calculating the integral $\mathcal{I}_2$ goes along the same lines. We finally obtain \begin{multline}\label{eq:integral1_sol}
% \begin{split}
  \mathcal{I} = -\frac{p^2}{(16\pi^2)^4} \\
\times \int_0^1 dx \int_0^{1-x}dy \int_0^1 dz \frac{xyz}{(1-z+zy^2-zy)^3} \left(\frac{2}{\epsilon} - \log \Delta_1 -\gamma + \log 4 \pi + \mathcal{O}(\epsilon) \right) \\
\times \int_0^1 da \int_0^{1-a}db \int_0^1 dc \frac{abc}{(1-c+cb^2-cb)^3} \left(\frac{2}{\epsilon} - \log \Delta_2 -\gamma + \log 4 \pi + \mathcal{O}(\epsilon) \right)\, . 
% \end{split}
\end{multline}
$\Delta_1$ is given in Eq.~\eqref{eq:delta1}, and for $\Delta_2$ we have
\begin{multline}
 \Delta_2 (a,b,c)= \left( \frac{abc}{(1-c+c b^2-cb)} p\right)^2 \\
- \frac{1}{(1-c+c b^2-cb)} \left[ca^2p^2-ca p^2+(1-a-b)c M_W^2-m_d^2\right]\, .
\end{multline}
In the case of $p^2=0$, these expressions simplify to
\begin{equation}
  \Delta_1(x,y,z) =  - \frac{1}{(1-z+zy^2-zy)} \left((1-x-y)zM_W^2-m_d^2\right)
\end{equation}
and
\begin{equation}
 \Delta_2(a,b,c) =  - \frac{1}{(1-c+cb^2-cb)} \left((1-a-b)c M_W^2-m_d^2\right)\, .
\end{equation}
The expression in Eq.~\eqref{eq:integral1_sol} can be renormalized via the usual ``minimal subtraction'' (MS) scheme.

We are now able to write down an expression for the correction to the neutrino mass generated by the diagram in Fig.~\ref{fig:blackboxmass}, assuming that there is no bare Majorana mass term in the Lagrangian. Plugging Eq.~\eqref{eq:integral1_sol} into Eq.~\eqref{eq:diagram1_intermed} we find
\begin{equation}
 \delta m_\nu = \Sigma(\Slash{p} = m_\nu) \propto \left. m_u^2 m_e^2 p^2\right|_{\Slash{p}=m_\nu}\, ,
 \label{eq:masscorrection0}
\end{equation}
where $m_\nu$ is the physical neutrino mass. We have neglected the dependence on $p^2$ inside the logarithms. As the physical neutrino mass $m_\nu$ is much smaller than $M_W$ this does not affect the discussion. The physical mass $m_\nu$ is the solution to the equation
\begin{equation}
 \left.\Slash{p} - \Sigma(\Slash{p})\right|_{\Slash{p}=m_\nu} = 0\, .
\end{equation}
One solution of this equation is $m_\nu=0$. The second solution $m_\nu = \alpha^{-1}$, where $\alpha$ is the proportionality factor omitted in Eq.~\eqref{eq:masscorrection0}, is huge due to the smallness of $\alpha$, and is unphysical as it does not correspond to any reasonable scale. Thus we obtain as physical correction to the zero neutrino mass:
\begin{equation}
 \delta m_\nu = 0\, .
\end{equation}
This means that the operator $J_L J_L j_L$ does not give a contribution via the Butterfly diagram shown in Fig.~\ref{fig:blackboxmass}. We checked the calculations numerically with the TARCER program~\cite{Mertig:1998vk}, which confirmed the zero result.

One may now wonder if there exists an operator generating a non-zero contribution via the Black Box diagram at all. It does exist, and we want to give an example for such a diagram in the next subsection. The result, however, will be many orders of magnitude smaller than the expected light neutrino mass.

Let us point out that the result of this subsection does not mean that all thinkable diagrams involving the operator $J_L J_L j_L$ will give a zero contribution like the one we calculated. Any other diagram, however, will include more loops and will therefore be suppressed more strongly. Thus, the mass we expect to find will be even smaller than the one we find in the next section for a different operator. However, one should keep the result of this subsection in mind when arguing for the Majorana nature of the electron neutrino via the diagram in Fig.~\ref{fig:blackboxmass}: Depending on the operator (i.e., on the underlying mechanism) of $0\nu\beta\beta$, one possibly draws a diagram leading to a zero mass. Clearly, this does not support the hypothesis of a Majorana neutrino (although there might be other diagrams giving a non-zero contribution for the same operator).

%%%%%%%%%%%%%%%%%%%%%%%%%%%%%%%%%%%%%%%%%%%%%%%%%%%%%%%%%%%%%%%%%%%%%%%%%
\subsection{Decay mediated by the Operator \texorpdfstring{$J^\mu_R J_{\mu R} j_L$}{JmuJmuj}}
%%%%%%%%%%%%%%%%%%%%%%%%%%%%%%%%%%%%%%%%%%%%%%%%%%%%%%%%%%%%%%%%%%%%%%%%%

In the previous subsection we have seen that the diagram calculated vanishes. However, there can of course be operators responsible for neutrinoless double beta decay which give non-zero contributions to the mass correction. In this subsection, we want to calculate the diagram assuming that the vertex is proportional to $\epsilon_3$ from Eq.~\eqref{eq:L_0nbb}. Additionally, we choose the chirality structure in the following way:
\begin{equation}
 J^\sigma_R J_{R\sigma} j_L = 8 \overline{u} \gamma^\sigma P_R d \, \overline{u}\gamma_\sigma P_R d \, \overline{e} P_L e^c\, .
\end{equation}
As in the former case, we have to find all contractions possible. So we can directly write down the matrix element of the diagram:
\begin{multline}\label{eq:diagram2}
%  \begin{split}
 \frac{8g^4 G_F^2 \epsilon_3}{ m_p} \int d\tilde{k}_1 d\tilde{k}_2 d\tilde{q}_1 d\tilde{q}_2 \frac{-i g_{\alpha\beta}}{(k_1^2-M_W^2)}\frac{-i g_{\mu\nu}}{(k_2^2-M_W^2)} \\ 
\mathrm{Tr} \left(i \frac{\Slash{k}_1+\Slash{q}_1+m_u}{(k_1+q_1)^2-m_u^2} \gamma^\sigma P_R i \frac{\Slash{q}_1+m_d}{q_1^2-m_d^2} \gamma^\beta P_L \right) \mathrm{Tr} \left(i \frac{\Slash{k}_2+\Slash{q}_2+m_u}{(k_2+q_2)^2-m_u^2} \gamma_\sigma P_R i \frac{\Slash{q}_2+m_d}{q_2^2-m_d^2} \gamma^\nu P_L \right) \\
\overline{\nu}(p) \gamma^\alpha P_L i \frac{\Slash{p}-\Slash{k}_2+m_e}{(p-k_2)^2-m_e^2} P_L i \frac{-\Slash{p}-\Slash{k}_1+m_e}{(p+k_1)^2-m_e^2} \gamma^\mu P_R \nu^c(p)\, .
% \end{split}
\end{multline}
Calculating the traces in a similar manner as before, we obtain
\begin{equation}
\mathrm{Tr} \left(i \frac{\Slash{k}_1+\Slash{q}_1+m_u}{(k_1+q_1)^2-m_u^2} \gamma^\sigma P_R i \frac{\Slash{q}_1+m_d}{q_1^2-m_d^2} \gamma^\beta P_L \right)
= - \frac{2 m_u m_d g^{\sigma\mu}}{[(k_1+q_1)^2-m_u^2][q_1^2-m_d^2]} 
\end{equation}
and
\begin{equation}
 \mathrm{Tr} \left(i \frac{\Slash{k}_2+\Slash{q}_2+m_u}{(k_2+q_2)^2-m_u^2} \gamma_\sigma P_R i \frac{\Slash{q}_2+m_d}{q_2^2-m_d^2} \gamma^\mu P_L \right)
= - \frac{2 m_u m_d g_\sigma^\mu}{[(k_2+q_2)^2-m_u^2][q_2^2-m_d^2]}\, .
\end{equation}
We can now rewrite the expression for the diagram as before:
 \begin{equation}
 \Sigma(p) = \frac{128 g^4 G_F^2 \epsilon_3 m_u^2 m_e^2 m_d^2 }{m_p}  \tilde{\mathcal{I}}\, ,
 \label{eq:diagram2_intermed}
\end{equation}
where the integral $\tilde{\mathcal{I}}$ is given by

\begin{equation}
 \tilde{\mathcal{I}} = \tilde{\mathcal{I}}_1 \times \tilde{\mathcal{I}}_2\, ,
\end{equation}

\noindent with

\begin{equation}
 \tilde{\mathcal{I}}_1 = \int d\tilde{k}_1 d\tilde{q}_1 \frac{1}{[(p+k_1)^2-m_e^2][(k_1+q_1)^2-m_u^2](q_1^2-m_d^2)(k_1^2-M_W^2)}
 \label{eq:integral1}
\end{equation}

\noindent and 

\begin{equation}
 \tilde{\mathcal{I}}_2 = \int d\tilde{k}_2 d\tilde{q}_2 \frac{1}{[(p-k_2)^2-m_e^2][(k_2+q_2)^2-m_u^2](q_2^2-m_d^2)(k_2^2-M_W^2)}\, .
 \label{eq:integral2}
\end{equation}

\noindent The factor $4$, which appears in Eq.~\eqref{eq:diagram2_intermed} but not in Eq.~\eqref{eq:diagram1_intermed}, arises from a contraction $\gamma^\mu\gamma_\mu=4$. The numerators in Eqs.~\eqref{eq:integral1} and~\eqref{eq:integral2} do not contain loop momenta, which is the crucial difference to the case discussed in the previous subsection.

To calculate $\tilde{\mathcal{I}}_1$ and $\tilde{\mathcal{I}}_2$, we can combine the factors in the denominators containing $k_1$ and $k_2$, respectively. Then, we may follow the calculations in~\cite{Ghinculov:1994sd}, such that we arrive at the expression
\begin{equation}
 \tilde{\mathcal{I}}_1 = \tilde{\mathcal{I}}_2 = \frac{1}{(4\pi^2)^{4+\epsilon}} \int_0^1 dz\ \mathcal{G} ( (1-z)^{1/2} M_W, m_d, m_u;0)\, ,
\end{equation}
with
\begin{multline}\label{eq:G}
 \mathcal{G} ( (1-z)^{1/2} M_W, m_d, m_u;0) = \\
 c_2(z) \epsilon^2 + c_1(z) \epsilon + \pi^4 \left\{ \frac{2}{\epsilon^2} + \frac{1}{\epsilon}\left[-1 + 2\gamma + 2 \ln(\pi(1-z) M_W^2 ) \right] + \frac{1}{4}+\frac{\pi^2}{12} \right. \\
 \left. +\frac{1}{4} \left[-1 + 2\gamma + 2 \ln(\pi(1-z) M_W^2 ) \right]^2-1+g( (1-z)^{1/2} M_W, m_d, m_u;0) \right\} \, ,
\end{multline}
where the functions $c_{1,2}$ are higher order corrections to the result from Ref.~\cite{Ghinculov:1994sd}, which we have computed numerically. Note that these corrections can normally be neglected when computing ordinary two-loop diagrams. Since we, however, effectively need a product of two such diagrams, the higher order terms can appear in products with terms of order $\mathcal{O}(\epsilon^{-1})$ or $\mathcal{O}(\epsilon^{-2})$, thereby leading to non-negligible finite corrections. The function $g$ appearing in Eq.~\eqref{eq:G} is defined by
\begin{multline}
 g( (1-z)^{1/2} M_W, m_d, m_u;0) = 1-\frac{1}{2}\ln a \ln b - \frac{a+b-1}{\sqrt{\Delta^\prime}} \\
 \times \left[ \mathrm{Sp}\left(-\frac{u_2}{v_1}\right) +\mathrm{Sp}\left(-\frac{v_2}{u_1}\right)+\frac{1}{4}\ln^2 \frac{u_2}{v_1}+\frac{1}{4}\ln^2 \frac{v_2}{u_1}+\frac{1}{4}\ln^2 \frac{u_1}{v_1} -\frac{1}{4}\ln^2 \frac{u_2}{v_1}+\frac{\pi^2}{6}\right]\, ,
\end{multline}
where
\begin{align}
 u_{1,2} &= \frac{1}{2} \left(1+b-a \pm\sqrt{\Delta^\prime}\right)\, ,\\
 v_{1,2} &= \frac{1}{2} \left(1-b+a \pm\sqrt{\Delta^\prime}\right)\, , \\
 \Delta^\prime &= 1-2(a+b) + (a-b)^2 \, ,
\end{align}
and
\begin{equation}
 a=\frac{m_d^2}{(1-z) M_W^2} , \quad  b=\frac{m_u^2}{(1-z) M_W^2}\, .
\end{equation}
The Spence function is given by
\begin{equation}
 \mathrm{Sp} (x) = \int_x^0 \frac{\ln(1-y)}{y} dy \, .
\end{equation}

Integrating over $z$ (the integration of $g$ and $c_{1,2}$ has been done
numerically), and expanding in $\epsilon$, we finally obtain
\begin{multline}\label{eq:expression_integrals}
  \tilde{\mathcal{I}}_1 = \tilde{\mathcal{I}}_2 = \frac{1}{(16\pi^2)^2} \Bigg[\frac{C_{-2}(M_W^2)}{\epsilon^2} +\frac{C_{-1}(M_W^2)}{\epsilon} \\ 
   + C_0 (M_W^2) + C_1(M_W^2) \epsilon + C_2(M_W^2) \epsilon^2 + \mathcal{O}(\epsilon^3) \Bigg]\, . 
\end{multline}

Here, the coefficients are given by
\begin{align}
 C_{-2}(x) &= 2, \\
 C_{-1} (x)&=   -3 + 2\gamma +2 \ln (x \pi) -2 \ln( 4\pi^2),\\
 C_0 (x)&= 2.64493 + \frac{1}{12}\left[30+12(-3+\gamma)\gamma+\pi^2 \right] + \ln (x\pi) \left[(-3+2\gamma)+ \ln(x\pi)\right] \nonumber \\
          &\quad + 2\ln^2( 4\pi^2) - \left[(-3+2\gamma) + 2 \ln (x\pi) \right] \ln( 4\pi^2),\\
 C_1 (x)&= \frac{A}{\pi^4} -\frac{\ln( 4 \pi^2)}{12} \left[30+12(-3+\gamma)\gamma+\pi^2 + 12 \ln ( x \pi) \left( (-3 +2\gamma) + \ln(x \pi)\right) \right] \nonumber \\ 
           &\quad -2.64493 \ln( 4 \pi^2) + \ln^2 (4 \pi^2) \left( -3+2\gamma + 2 \ln(x\pi) \right) -2 \ln^3( 4 \pi^2),
\end{align}
\begin{align}
 C_2 (x) &= \frac{B}{\pi^4}- \frac{A}{\pi^4}\ln (4 \pi^2)  \nonumber \\
   &\quad + \frac{1}{12}\left[30+12(-3+\gamma)\gamma+\pi^2 + 12 \ln ( x \pi) \left( (-3 +2\gamma) + \ln ( x \pi)\right) \right] \ln^2 ( 4 \pi^2) \nonumber \\ 
  &\quad + 2.64493  \ln^2 ( 4 \pi^2) - \ln^3 ( 4 \pi^2) \left[(-3+2\gamma) + 2 \ln(x\pi) \right] + 2 \ln^4 ( 4 \pi^2), \, 
\end{align}
and
\begin{align}
 A &= \int_0^1 dz \, c_1(z) = 3.93180\times 10^5\, , \\
 B &= \int_0^1 dz \, c_2(z) = 2.28888\times 10^6\, .
\end{align}
Then,
\begin{multline}\label{eq:expression_integrals2}
\tilde{\mathcal{I}} = \tilde{\mathcal{I}}_1 \times \tilde{\mathcal{I}}_2 = \frac{1}{(16\pi^2)^4} \Bigg\{ \frac{C_{-2}^2(M_W^2)}{\epsilon^4} + \frac{2 C_{-2}(M_W^2) C_{-1}(M_W^2)}{\epsilon^3}  \\
+ \frac{2 C_0 (M_W^2) C_{-2} (M_W^2) + C_{-1}^2 (M_W^2)}{\epsilon^2} \\
+ \frac{2 C_{-2}(M_W^2)C_1(M_W^2) + 2 C_{-1}(M_W^2) C_0(M_W^2)}{\epsilon}\\
 + C_0^2(M_W^2) +2 C_{-1}(M_W^2) C_1(M_W^2) + 2 C_{-2}(M_W^2)C_2(M_W^2) + \mathcal{O}(\epsilon) \Bigg\} \\
 \xrightarrow{\mathrm{MS}}\frac{1}{(16\pi^2)^4} \left[C_0^2(M_W^2/\mu^2) + 2 C_{-1}(M_W^2/ \mu^2) C_1(M_W^2/ \mu^2) \right. \\
\left. + 2 C_{-2}(M_W^2/\mu^2)C_2(M_W^2/ \mu^2) \right] \, .
\end{multline}

As we are interested only in an order of magnitude estimate for the neutrino mass correction, we can renormalize the expression in Eq.~\eqref{eq:expression_integrals2} via the $\mathrm{MS}$ scheme (which means that we subtract the poles in $\epsilon$) and take the constant term as the leading contribution. Thus we find for the mass corrections
\begin{multline}\label{eq:masscorrection_final}
 \delta m_\nu = \frac{128 g^4 G_F^2 \epsilon_3 m_u^2 m_e^2 m_d^2 }{(16\pi^2)^4 m_p} \\
\times \left[C_0^2(M_W^2/\mu^2) + 2 C_{-1}(M_W^2/ \mu^2) C_1(M_W^2/ \mu^2) + 2 C_{-2}(M_W^2/\mu^2)C_2(M_W^2/ \mu^2) \right] \, .
\end{multline}
Using a renormalization scale $\mu = 100\, \mathrm{MeV}$, as appropriate for a nuclear physics problem, we obtain\footnote{Earlier versions of this paper (including the version published in JHEP) state a value of $\delta m_\nu = 9.4\times 10^{-25}\, \mathrm{eV}$, due to a numerical typo in the Mathematica script. We thank Martin Hirsch and Jun-Hao Liu for pointing us to this issue. Note, however, that the exact numerical value does not affect our conclusions. }
\begin{equation}
  \delta m_\nu = 5\times 10^{-28} \, \mathrm{eV} \, .
\end{equation}

Here, we have used the values for masses and constants given in Tab.~\ref{tab:constants}. It is obvious that this correction is far too small to be the main contribution to the neutrino masses we expect to have. In particular, such a mass cannot explain the neutrino oscillation data. Therefore, another mechanism must give the leading contribution of neutrino masses. However, if we know that a Majorana neutrino mass is generated in a New Physics model anyway, we do not need the Butterfly diagram anymore.

\begin{table}
\centering
 \begin{tabular}{c@{\hspace{1ex}}c@{\hspace{1ex}}lc@{\hspace{1ex}}c@{\hspace{1ex}}l}\hline
  $m_u$&$=$&$1.5\text{~to~}3.3\,\mathrm{MeV}$ & $m_d$&$=$&$3.5\text{~to~}6.0\,\mathrm{MeV}$\\
  $m_e$&$=$&$(0.510998910\pm 0.000000013)\,\mathrm{MeV}$ & $m_p$&$=$&$(938.27203\pm0.00008)\,\mathrm{MeV}$ \\
  $G_F$&$=$&$1.16637(1)\times 10^{-5}\,\mathrm{GeV}^{-2}$ & $g$&$=$&$0.652$\\ 
  $M_W$&$=$&$(80.399 \pm 0.023)\,\mathrm{GeV} $ & & & \\ \hline
 \end{tabular}
\caption{\label{tab:constants} Masses and constants used in the calculation of Eq.~\eqref{eq:masscorrection_final} (taken from~\cite{Amsler:2008zzb}).}
\end{table}

%%%%%%%%%%%%%%%%%%%%%%%%%%%%%%%%%%%%%%%%%%%%%%%%%%%%%%%%%%%%%%%%%%%%%%%%%
\section{\label{sec:chiral} The Relation to Chiral Symmetry}
%%%%%%%%%%%%%%%%%%%%%%%%%%%%%%%%%%%%%%%%%%%%%%%%%%%%%%%%%%%%%%%%%%%%%%%%%

Finally, we want to consider the question in how far non-zero contributions to neutrino masses are related to the breaking of chiral symmetry. For example, in QCD it is well known that fermion mass terms necessarily break chiral symmetry, as they connect left-handed fermions with right-handed ones~\cite{1995nucl.th..12029K}. Hence, it is natural to ask the question whether the operators responsible for $0\nu\beta\beta$ violate chiral symmetry, as necessary for an operator that finally leads to a neutrino mass.

Chiral symmetry acts on fermion fields as
\begin{equation}
 \psi(x) \rightarrow e^{i\alpha\gamma_5} \psi(x)\simeq \left(1+ i\alpha\gamma_5\right) \psi(x)\, ,
 \label{eq:chiraltrafo}
\end{equation}
where the last step is valid for infinitesimal transformations. In this limit, it is easy to show that the usual right-handed Majorana mass term, which looks like
\begin{equation}
 \mathcal{L}_{\rm Majorana}=-M_R \overline{\psi^c} P_R \psi
 \label{eq:Majorana_4spinor}
\end{equation}
in this notation, does not conserve chiral symmetry, as to be expected. The behavior of the different operators from Eqs.~\eqref{eq:operators_1} and~\eqref{eq:operators_2} is summarized in Tab.~\ref{tab:symmop}. Obviously, the only operators that are invariant under chiral symmetry are the ones with exactly one Lorentz index. Glancing at Eq.~\eqref{eq:L_0nbb} we can easily see that, in order to transmit $0\nu\beta\beta$, the combinations of operators must be of the form ``$JJj$,'' which is required by the simple fact that we need exactly two outgoing electrons. Such combinations of three currents can, however, never be built out of three operators with exactly one Lorentz index each, since a full contraction of the indices would never be possible. Accordingly, any operator that could be responsible for $0\nu\beta\beta$ must necessarily violate chiral symmetry. Turning this round, we have no operator in Eq.~\eqref{eq:L_0nbb} that would be forbidden to yield a finite neutrino mass.

\begin{table}
\centering
 \begin{tabular}{ll}\hline
  Operator & Chiral Symmetry \\ \hline \hline 
  $\overline{e}\left( 1 \mp \gamma_5 \right) e^c$ & not invariant\\ 
  $\overline{e} \gamma^\mu\left( 1 \mp \gamma_5 \right) e^c$ & invariant  \\ 
  $\overline{u}\left( 1 \mp \gamma_5 \right) d$ & not invariant \\ 
  $\overline{u}\gamma^\mu\left( 1  \mp \gamma_5 \right) d$ & invariant \\ 
  $\overline{u}\frac{i}{2}[\gamma^\mu,\gamma^\nu]\left( 1 \mp \gamma_5 \right) d$ & not invariant \\ \hline
 \end{tabular}
\caption{\label{tab:symmop} Behavior of the different operators under chiral symmetry.}
\end{table}

So why do not all operators lead to a non-zero correction? The simple answer to this question is that all of them will lead to some non-zero correction, but not necessarily at the order of the Butterfly diagram. The breaking of chiral symmetry is a necessary, but not a sufficient condition for the existence of the particular mass contribution under consideration. Accordingly, the Schechter-Valle theorem remains valid, with the only addendum that in some cases it might be necessary to draw a diagram that is more complicated than the one given in Fig.~\ref{fig:blackboxtheorem}.

%%%%%%%%%%%%%%%%%%%%%%%%%%%%%%%%%%%%%%%%%%%%%%%%%%%%%%%%%%%%%%%%%%%%%%%%%
\section{\label{sec:conc}Conclusions}
%%%%%%%%%%%%%%%%%%%%%%%%%%%%%%%%%%%%%%%%%%%%%%%%%%%%%%%%%%%%%%%%%%%%%%%%%

We have discussed in this paper the relation between the so-called Black Box operator(s), which are responsible for neutrinoless double beta decay, and neutrino mass operators. We emphasized that one has to be careful relating a possible observation of neutrinoless double beta decay fully to neutrino masses, while a very tiny Majorana contribution to the neutrino mass is guaranteed. The Black Box operator generates a Majorana mass contribution via the diagram in Fig.~\ref{fig:blackboxmass}, which makes neutrinos Majorana particles, but we showed that the numerical contribution is many orders of magnitude too small to account for the observed neutrino masses. We have even identified one operator mediating $0\nu\beta\beta$, but giving a zero contribution to the neutrino mass at leading (four loop) order in perturbation theory, demonstrating that the true Majorana contribution could even be further suppressed. We conclude therefore that lepton number violating operators beyond the Standard Model, which are not directly related to neutrino masses, guarantee radiatively non-zero Majorana neutrino mass terms, but the numerical values are extremely tiny such that other leading contributions to the physical neutrino mass must necessarily exist. If these mass terms are of Majorana nature, then they will dominate the Black Box diagram and there exists a direct relation to the observed rate of neutrinoless double beta decay. The Black Box induced Majorana mass is in this case only an extremely tiny correction found to be $\delta m_\nu = \mathcal{O}(10^{-28} \, \mathrm{eV})$ which can for practical purposes be safely ignored. If, on the other hand, neutrinos are predominantly Dirac particles, then neutrinoless double beta decay is still possible due to other lepton number violating operators. The Black Box will then induce a tiny Majorana contribution, making neutrinos Majorana particles, but the numerical value is so tiny that one might consider this more or less as an academic statement. The important point is that neutrinoless double beta decay might exist as consequence of other lepton number violating physics beyond the Standard Model which is essentially unrelated to the values of the neutrino masses (an example for such a model can be found in Ref.~\cite{Bhattacharyya:2002vf}). Let us rephrase this statement, to make it even clearer: Of course, if the neutrino is a Majorana particle and if we find that $0\nu\beta\beta$ is predominantly mediated by light neutrino exchange, the experiments are able to test the neutrino mass scale. However, let us point out that we will not be able to find out if this is the case by experiments on neutrinoless double beta decay alone. This should be kept in mind when possible signals for neutrinoless double beta decay are discussed and translated into neutrino masses.

Finally we would like to emphasize that our findings are logically fully consistent with the Schechter-Valle theorem, as the proof presented in Sec.~\ref{sec:bbt} excludes that a Majorana mass term is forbidden by symmetry provided that neutrinoless double beta decay takes place. We have discussed how the well-known diagram may vanish at leading order, but at higher order there must exist a Majorana contribution.

%%%%%%%%%%%%%%%%%%%%%%%%%%%%%%%%%%%%%%%%%%
\section*{Acknowledgments}
%%%%%%%%%%%%%%%%%%%%%%%%%%%%%%%%%%%%%%%%%%

We are grateful to Fedor \v{S}imkovic for kindly providing us with his values obtained for the nuclear matrix elements. This work has been supported by the DFG Sonderforschungsbereich Transregio 27 ``Neutrinos and beyond -- Weakly interacting particles in Physics, Astrophysics and Cosmology''. MD is supported by the International Max Planck Research School (IMPRS) ``Precision Tests of Fundamental Symmetries'' of the Max Planck Society. The work of AM is supported by the Royal Institute of Technology (KTH), under project no. SII-56510.

\begin{small}
\bibliographystyle{./utcaps_mod}
\bibliography{./sv_theorem.bib}
\end{small}

\end{document}